\documentclass[12pt,preprint]{aastex}

\shorttitle{Kitt Peak Sky Brightness}

\shortauthors{Neugent et al.}

\begin{document}

\title{The Spectrum of the Night Sky Over Kitt Peak:\\ Changes Over Two Decades}

\author{Kathryn F. Neugent\altaffilmark{1} and Philip Massey\altaffilmark{1}}
\affil{Lowell Observatory, 1400 W Mars Hill Road, Flagstaff, AZ 86001; \\kneugent@lowell.edu; phil.massey@lowell.edu}

\altaffiltext{1}{Visiting Astronomer, Kitt Peak National Observatory, National Optical Astronomy Observatory, which is operated by the Association of Universities for Research in Astronomy, Inc.\ (AURA) under cooperative agreement with the National Science Foundation.}


\begin{abstract}
New absolute spectrophotometry of the Kitt Peak night sky has been obtained in 2009/10, which we compare to previously published data obtained in 1988 and 1999, allowing us to look for changes over the past two decades. A comparison of the data between 1988, 1999 and 2009/10 reveals that the sky brightness of Kitt Peak has stayed remarkably constant over the past 20 years. Compared to 1988, the 2009/10 data show no change in the sky brightness at Zenith though, as expected, the sky glow has increased most dramatically in the direction of Tucson. Comparisons between the 1999 and 2009/10 data suggest that the sky has actually $\emph{decreased}$ in brightness compared to 10 years ago. However, the older data were both taken during times of increased solar activity. When we correct the measurements for the solar irradiance fluctuations, we find that compared to 20 years ago, the sky is $\sim$0.1 magnitude brighter at Zenith and $\sim$0.3 magnitudes brighter towards Tucson. But even after these corrections, we still find that the sky over Kitt Peak is comparable to what it was 10 years ago at Zenith and $\sim$0.1 magnitude darker towards Tucson. This suggests that the strengthened lighting ordinances Tucson and Pima County established in the early 2000s have been quite effective. With some care, the Kitt Peak night sky will remain this dark many years into the future. 
\end{abstract}

\section{Introduction}
\label{INTRO}
\begin{quotation}
``The present sky glow from Tucson is appreciable on the NE horizon ...\ the Glow does not extend very high in the sky ...\ nevertheless the situation could become an annoyance if Tucson should expand westward. The availability of ...\ water in the indefinite future could change the entire face of Arizona, but the development of a space observatory may have rendered this eventuality of minor consequence to the future of astronomy by that time." Aden Meinel, 1958
\end{quotation}

Over 50 years ago, Aden Meinel selected Kitt Peak as a promising location for the national observatory because of its high elevation, low humidity and pristine dark skies. But contrary to Meinel's prediction, the lack of water in Tucson has done little to slow its growth. Instead, the population of neighboring Pima County, which contains Tucson, has nearly quadrupled (U.S. Census Bureau: 1960, 2010\footnote{http://www.census.gov}) since Meinel spoke these words. Previously, Massey et al.\ (1990) and Massey \& Foltz (2000) discovered that this population increase, and that of other neighboring towns, has had little affect on the Kitt Peak sky brightness. Our research expands upon these studies by revisiting this question a full decade later.

To make this determination, we used spectrophotometry rather than broadband measurements because spectrophotometry clearly shows the relative contribution of atmospheric and human-caused light sources. In contrast, broadband measurements only show the overall brightness caused by all sources. Thus, it is very difficult to determine which specific lines are major contributors to the overall sky brightness. Our spectrophotometry focuses on the blue and yellow portions of the spectrum and puts less emphasis on the red, which is heavily dominated by OH emission intrinsic to the Earth's atmosphere.

\section{Observations and Reductions}
\label{OS}
\begin{figure}
\epsscale{1.0}
\plotone{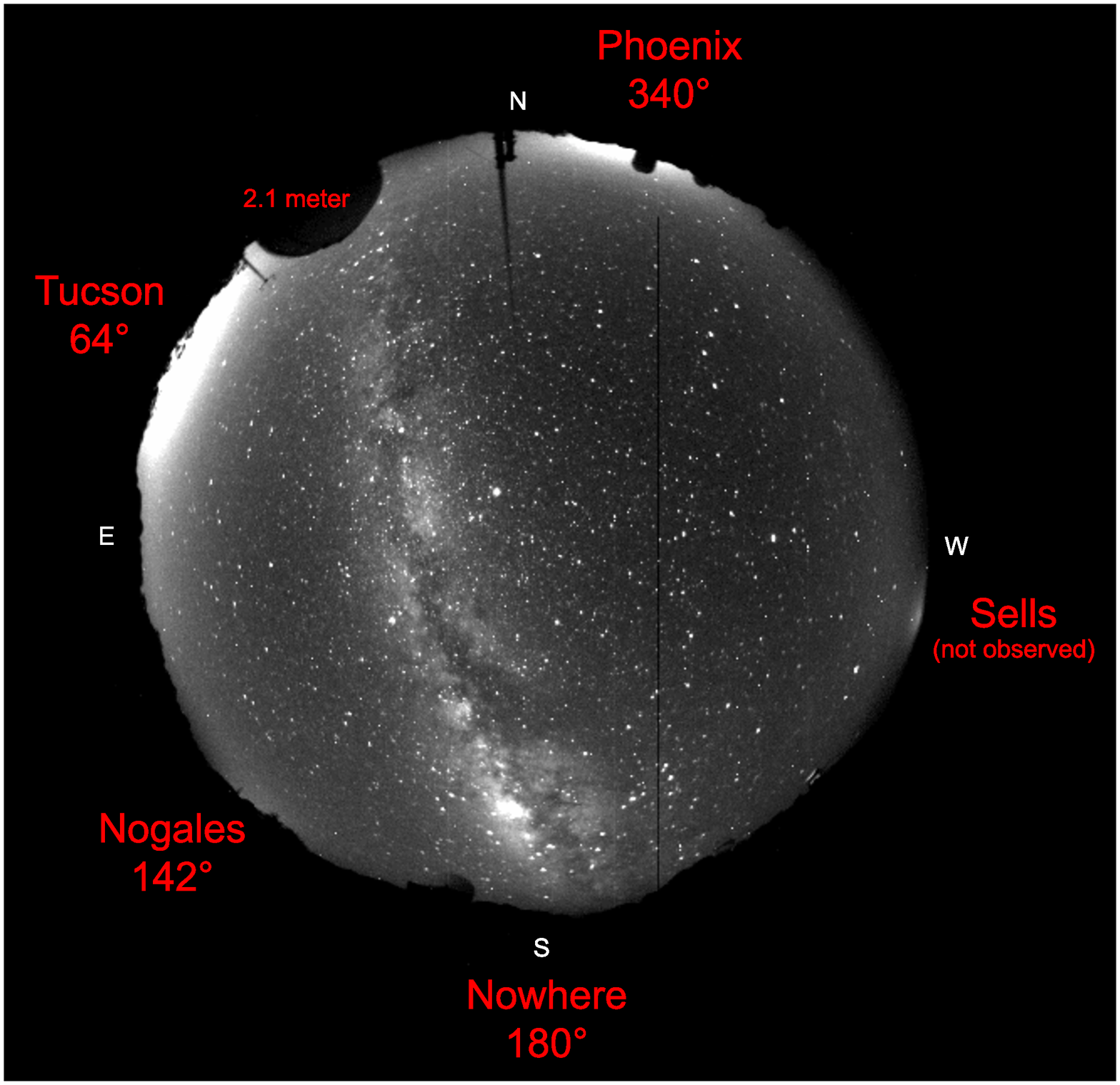}
\caption{\label{fig:locations} Sky Glow. This Kitt Peak All-Sky Camera image was taken on 2010 June 13.3 in the $B$ filter and shows the locations and relative sky glow caused by Tucson, Phoenix, Nogales, Sells and Nowhere. Note that observations were taken at a zenith distance of 60$^\circ$ and at least $15^\circ$ away from the ecliptic and the Galactic Plane. The labels simply denote the directions of the cities.}
\end{figure}

We strove to keep our observations as consistent with the 1988 and 1999 procedures (described in Massey et al.\ 1990 and Massey \& Foltz 2000, respectfully) as the changing equipment allowed. The 2009/10 data were taken on the Kitt Peak 2.1 meter telescope using the GoldCam CCD spectrometer during prime dark time. All observations were taken at least $15^\circ$ away from the ecliptic and the Galactic Plane, in order to avoid contributions from Zodiacal Light and the Milky Way. Measurements were made either at Zenith, or Zenith distances of $60^\circ$ towards Tucson (azumuth = 64$^\circ$), Phoenix (azimuth = 340$^\circ$), Nogales (azimuth = 142$^\circ$) or ``Nowhere" (azimuth = 180$^\circ$) as shown in Figure~\ref{fig:locations}. A 600 mm$^{-1}$ grating (``No.\ 26 new") was used in first order to provide a wavelength coverage of 3700 \AA\ -- 6800 \AA\ (after the data were trimmed), with a dispersion of 1.24 \AA\ per pixel. The data were binned by 3 pixels in the spatial direction to reduce noise (2.3$\arcsec$ binned pixels). The wide slit resulted in a spectral resolution of 6.6 \AA\ (5.3 pixels), comparable to that used in our earlier studies (Massey et al.\ 1990, Massey \& Foltz 2000). Normally, a 150$\arcsec$ metal decker plate was used to reduce scattered light, although for the October 2009 observations, a 100$\arcsec$ decker was (accidentally) used.

All blank sky observations were taken in sets of three 15 minute exposures with the telescope tracking on either (UT) 24 March 2009, 17 October 2009, 14 February 2010, 15 February 2010, 13 June 2010 or 14 June 2010. Between each exposure we reset the telescope back to the desired Zenith distance and azimuth. Based upon a visual inspection of the Kitt Peak National Observatory all-sky camera\footnote{http://www-kpno.kpno.noao.edu/Info/Mtn\_Weather/allsky/} and the sky, these appeared to be all-sky photometric nights. We additionally observed $\sim$10 spectrophotometric standards for flux calibration purposes each night over a range of airmass (Massey et al.\ 1988).

The most difficult step when determining the sky brightness spectrophotometrically is to determine the plate scales, at the entrance slit and detector, accurately enough for precise ($<$ 0.05 magnitude) measurements. For the plate scale at the slit, we adopted the canonical 12.7$\arcsec$ mm$^{-1}$ scale found in the observing manual. This is consistent with our own measurements of the scale from images taken with the direct CCD camera, which yields 0.305$\arcsec$ per 24 $\mu$m pixel on the direct CCD. Thus, we consider the scale at the slit to be determined to better than 1\%. We directly measured the scale at the detector using UCAC2 stars in M67 and obtained a scale of $0.765\arcsec \pm 0.003\arcsec$ per pixel based upon repeated measurements. So, this value is also known to much better than 1\%.

The data were reduced in a manner similar to that of Massey \& Foltz (2000). They were first corrected for the bias level by removing the average value in the overscan strip on a row-by-row basis. This removed some otherwise troublesome banding present in the data. We examined an average bias frame and found no residual counts or structure greater than 1 electron. The data were trimmed to 2531 pixels in the wavelength direction. Data outside of this region had poor focus or too few counts per exposure in the UV. We trimmed the data to either 45 or 70 binned pixels in the spatial direction depending on which decker was used. A bad pixel mask was constructed from the division of long and short exposures of an internal quartz lamp, and was used to remove the few bad columns and occasional bad pixel by linear interpolation. Dome flats were used to correct for pixel to pixel sensitivity variations, the response of the grating in the wavelength direction, and first-order vignetting in the spatial direction. We then fit a function to the twilight flats to further correct for the slit illumination function which, on average, represented around a 3\% change. Additionally, we created a ``master" sky image by combining the blank sky images for each night. A low order function was then fit to the spatial variance of this image and all of the images were then divided by this fit as a further spatial correction, typically $<$1\%. Finally, HeNeAr lamp exposures were used to provide wavelength calibrations. 

Since it was vital that our ``blank sky" images be completely devoid of stars, we strove to position our slit as far away from surrounding stars as possible. However, very dim stars did occasionally make their way into one of our frames. Thus, we combined our three consecutive images using a discrepant pixel rejection algorithm (``avsigclip") using IRAF's\footnote{IRAF is distributed by National Optical Astronomy Observatory, which is operated by the Association of Universities for Research in Astronomy, Inc.\ (AURA) under cooperative agreement with the National Science Foundation.} ``imcombine" routine. This removed cosmic rays and the spectra of any faint, resolved star.

The standard star observations were extracted using a moderately wide extraction aperture, and we traced the spectrum along the detector. The same trace was used for the comparison exposure in order to provide a wavelength scale. The standards were taken from Massey et al.\ (1988), and we constructed a wavelength-dependent sensitivity curve using these data. The scatter of our (typically 10) standards was $\sim$0.03 magnitudes. If the observations from two consecutive nights were involved, we combined the data using a common flat field. An average extinction curve was used to reduce the data. For the sky brightness data we used an extraction aperture of either 40 pixels (30.6$\arcsec$) or 60 pixels (45.9$\arcsec$) depending upon the decker used. 

Errors for our data can be estimated as the quadrature sum of the individual uncertainties. The plate scale is well determined ($<$1\%), as described previously. Nevertheless, in setting the slit width to 395 $\mu$m (5.00$\arcsec$) we relied upon our ability to read the micrometer setting by eye, a process which was reproducible only at the 3 $\mu$m level. In addition, there was the question of whether the slit was exactly fully closed when the micrometer read 0. We measured this repeatedly using exposures of an internal quartz lamp, varying the slit width, and found that the zero-point was no larger than 3 $\mu$m. Thus, our slit setting is probably uncertain by about 4 $\mu$m, or 2\% (0.02 magnitudes). As described previously, the uncertainty in the plate scale at the detector is known to better than 1\%. The adoption of a mean extinction curve was estimated by Massey \& Foltz (2000) to add another 0.01~magnitude error, since this affects only the wavelength dependent part of the extinction. The mean errors from the spectrophotometric standards adds another 0.03~magnitudes, a combination both of the uncertainty in the absolute calibration and probably variable slit losses. So in total, we estimate the uncertainty in our spectrophotometry as 0.04~magnitudes.

\section{Results}
\label{R}
Although our goal was to obtain moderate resolution spectrophotometry, it is also useful to describe the data in terms of broad- and narrow-band indices, to facilitate comparison with the previous two decades. The synthetic $V$ and $B$ magnitudes presented in Tables~\ref{tab:AllMags} and \ref{tab:CompMags} were computed using the filter functions determined by Bessell (1990). The convolutions for both the broadband and narrowband magnitudes were performed on Vega using STIS spectrophotometry (Bohlin \& Gilliland 2004) to determine the zero-points assuming that Vega has a $B$ and $V$ magnitude of 0.03 (Bessell et al.\ 1988). The broadband magnitudes were then computed after replacing 17 \AA\ on both side of the highly variable OI 5577 atmospheric line by an averaged value. The narrowband magnitudes were computed using a constant response over a 100 \AA\ interval centered at either $\lambda$4250, $\lambda$4550, or $\lambda$5150. The 1988 and 1999 magnitudes presented in Tables~\ref{tab:AllMags} and \ref{tab:CompMags} do not perfectly match those of Massey et al.\ (1990) and Massey \& Foltz (2000), as there was an error made in the conversion to the standard system in the previous studies. The values given here are correct, and agree much better with other broadband measurements of the night sky brightness of Kitt Peak, e.g., Pilachowski et al.\ (1989). 

\begin{figure}
\epsscale{1.0}
\plotone{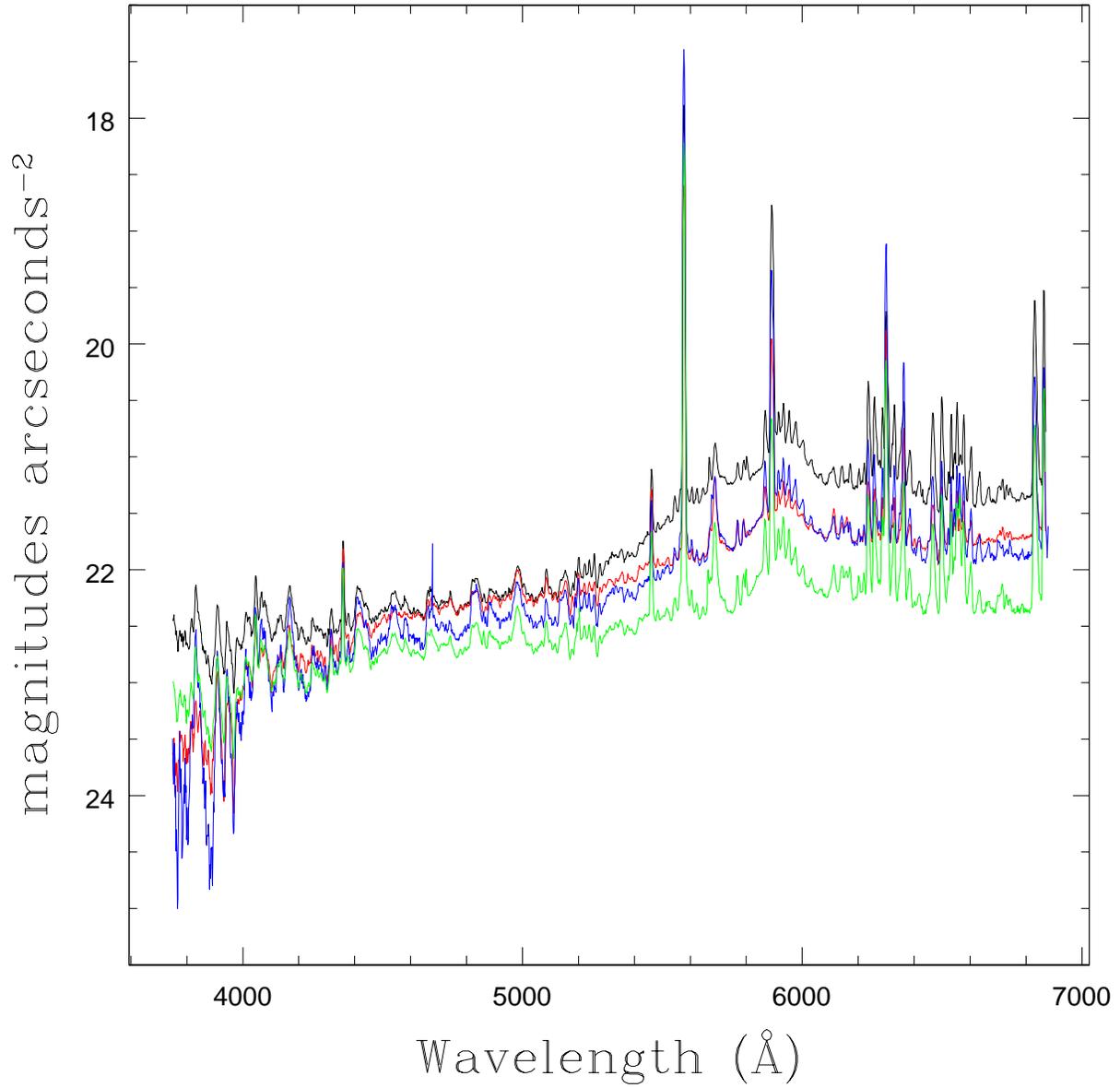}
\caption{\label{fig:Octdiff} Differences in 2009/10 Zenith spectra. This shows the differences at Zenith between the February data (red), the March data (blue), the October data (black) and the June data (green). While the February and the March data are almost identical, both the October and June data display strong differences in the red portion of the spectrum.}
\end{figure}

Since all of our 2009/10 data were taken on all-sky photometric nights, we were especially interested in any night-to-night variations over our 15 month span of observations. Figure~\ref{fig:Octdiff} presents a comparison in the 2009/10 Zenith data. While the 2010 February and 2009 March Zenith observations are almost indistinguishable, the 2009 October and 2010 June data both differ by close to a half a magnitude per arcsec$^{2}$ in the red. For the October data, this trend was also observed when looking at Tucson, Nogales, Phoenix and ``Nowhere" and the broadband $V$ and $B$ magnitudes in Table~\ref{tab:AllMags} also show that the October magnitudes are consistently half a magnitude brighter. However, for the June data, dimming in the red was only observed in the Zenith measurements. We haven't found any acceptable explanations for this behavior. Though, a similar sky brightness study at Mt. Graham found that non-photometric conditions can cause a 0.5 magnitude increase in sky brightness, consistent with our October results (Pedani 2009). Thus, when an average spectrum or magnitude is shown (such as in Table~\ref{tab:CompMags}), we have combined only the February, March and June data except for at Zenith, where only the February and March data have been combined. 

\begin{figure}
\epsscale{0.52}
\plotone{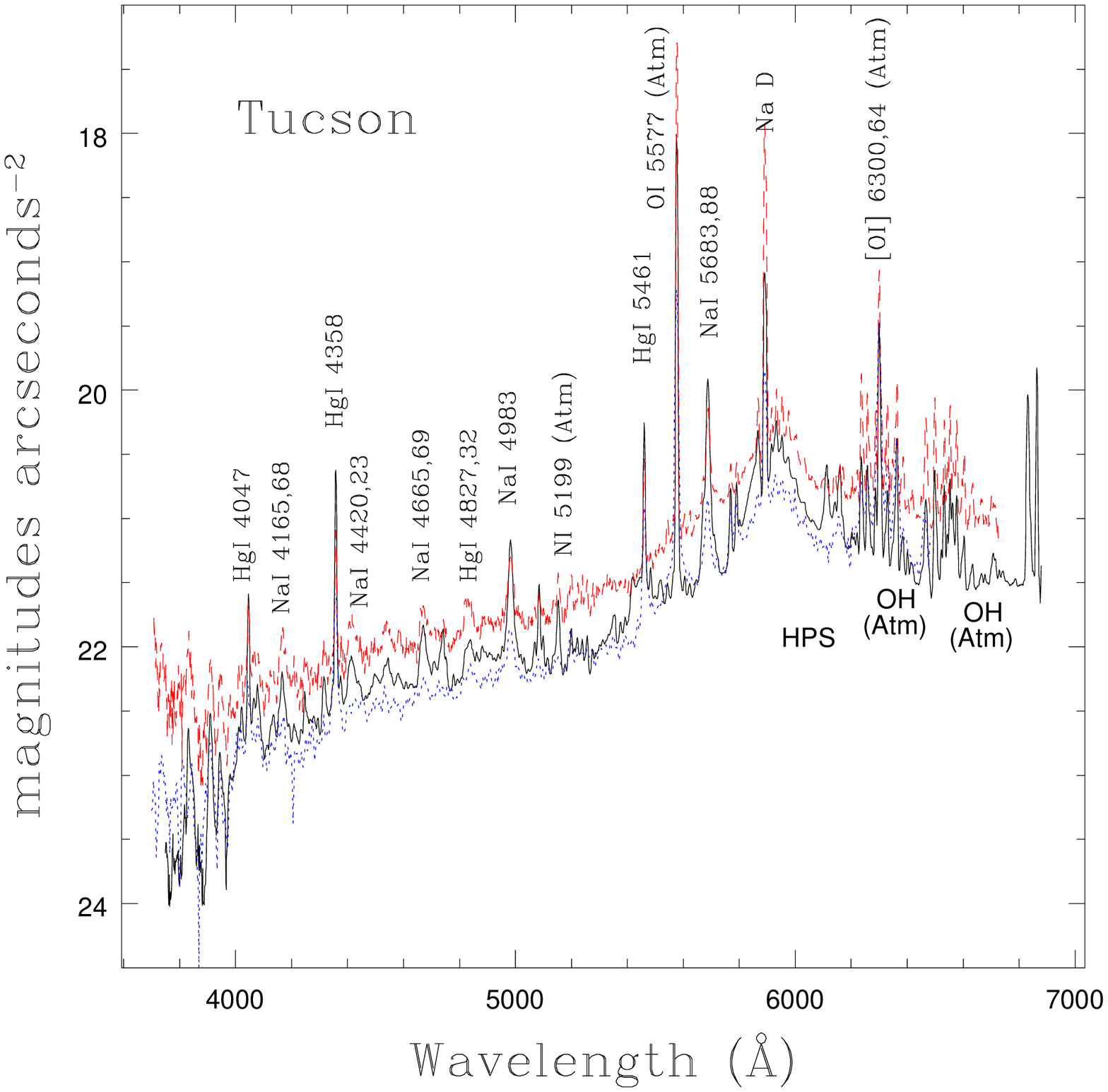}
\plotone{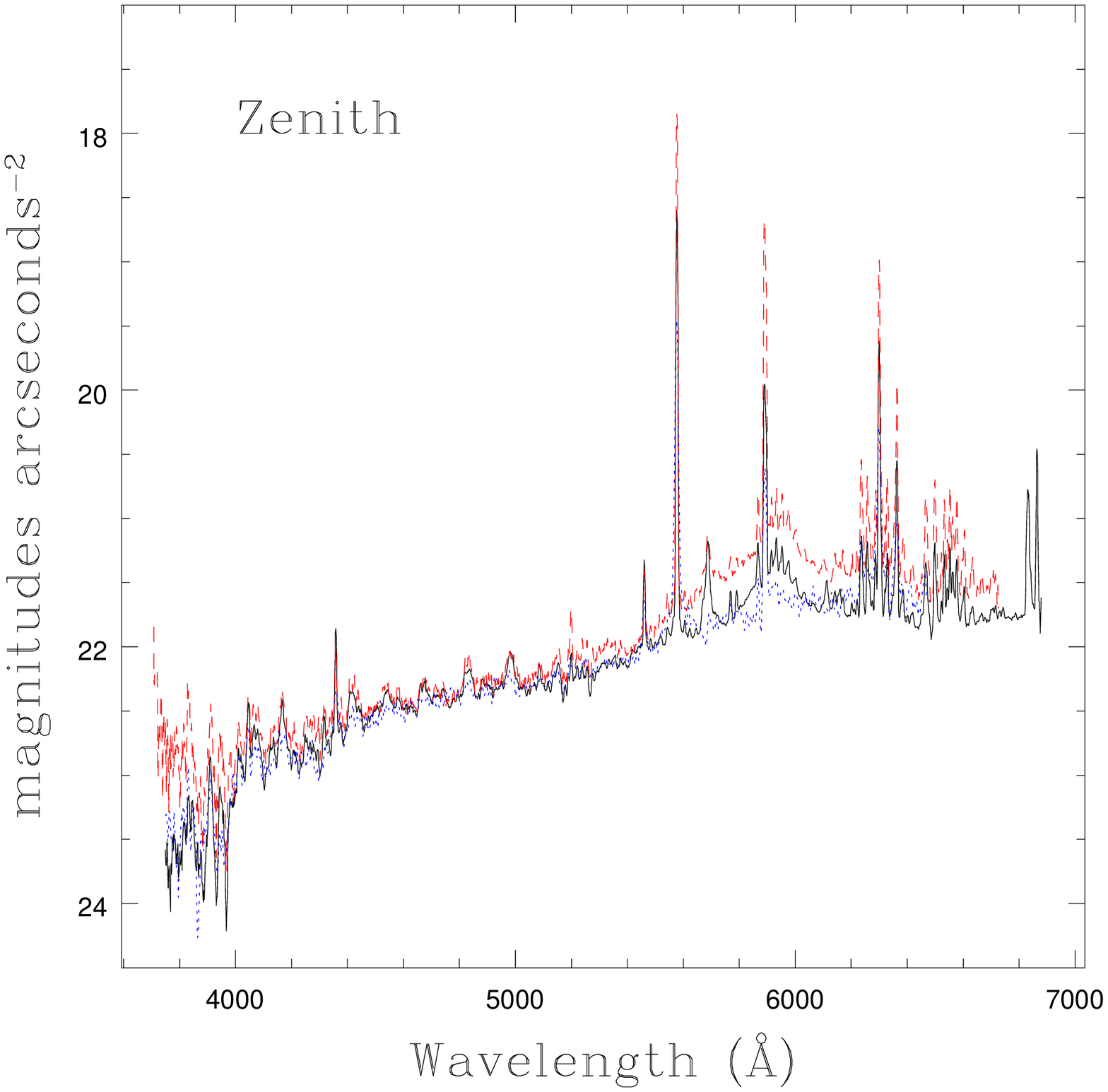}
\caption{\label{fig:20yrs} Brightness changes over the past 20 years. The top figure shows the changes in the Tucson spectrum with the prominent atmospheric and artificial lines labelled while the bottom figure shows the changes in the Zenith spectrum. In both plots, the black solid line represents the 2009/10 data, red dashed line represents the 1999 data and the blue dotted line represents the 1988 data.}
\end{figure}

Arguably the most interesting decade to decade comparisons deal with the sky brightness at Tucson and Zenith. Figure~\ref{fig:20yrs} shows that in both cases the largest change in brightness occurred between 1988 and 1999 and after 1999, the sky appears to have gotten \emph{darker}. A magnitude comparison between the 1999 and 2009/10 values in Table~\ref{tab:CompMags} shows that both the broadband and narrowband magnitudes additionally decreased. Much of this change may be caused by the solar cycle as well as neighboring Pima County's stringent lightening ordinances put into place in 2000, as discussed in Section~\ref{D}.

\begin{figure}
\epsscale{0.48}
\plotone{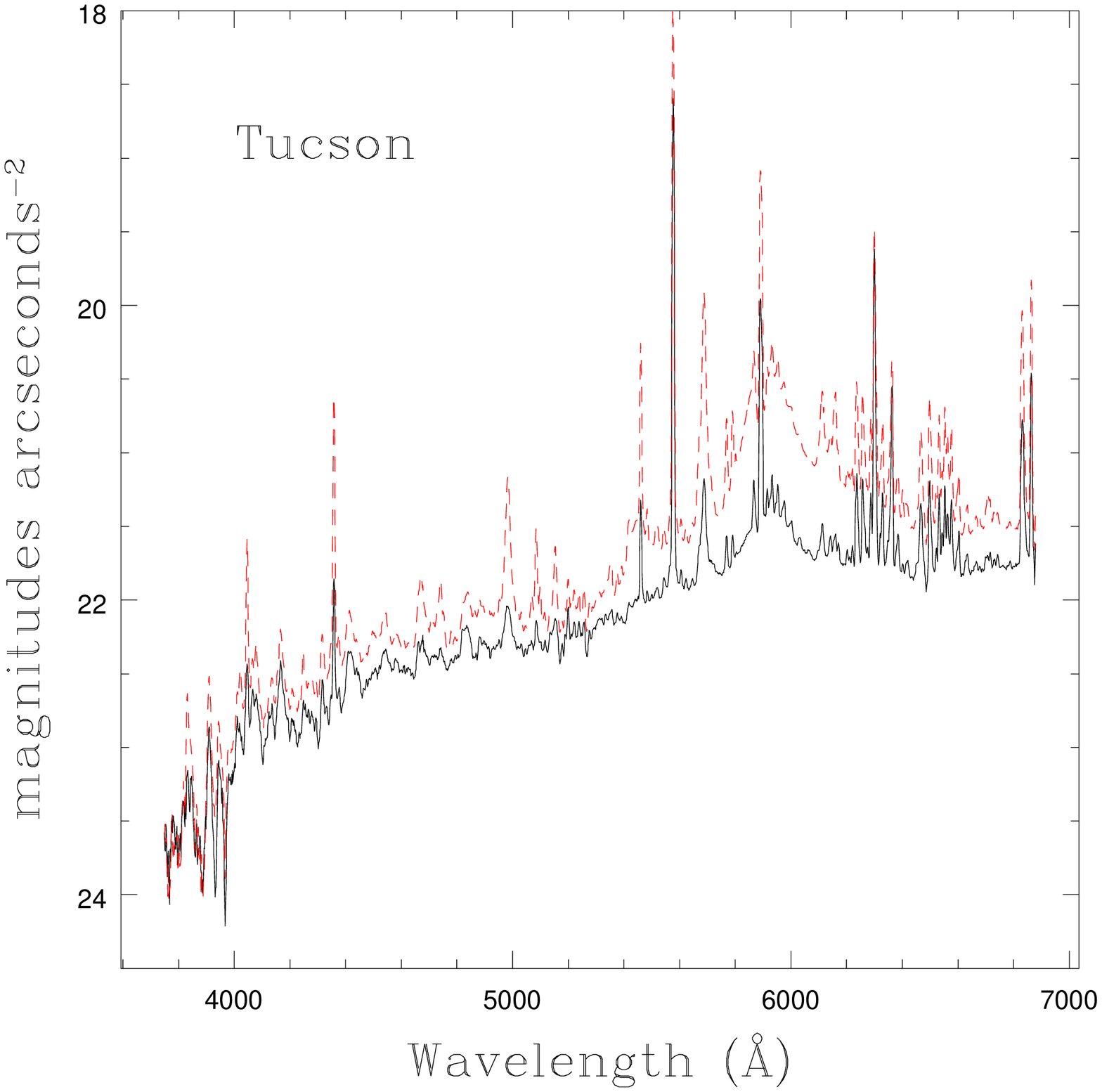}
\plotone{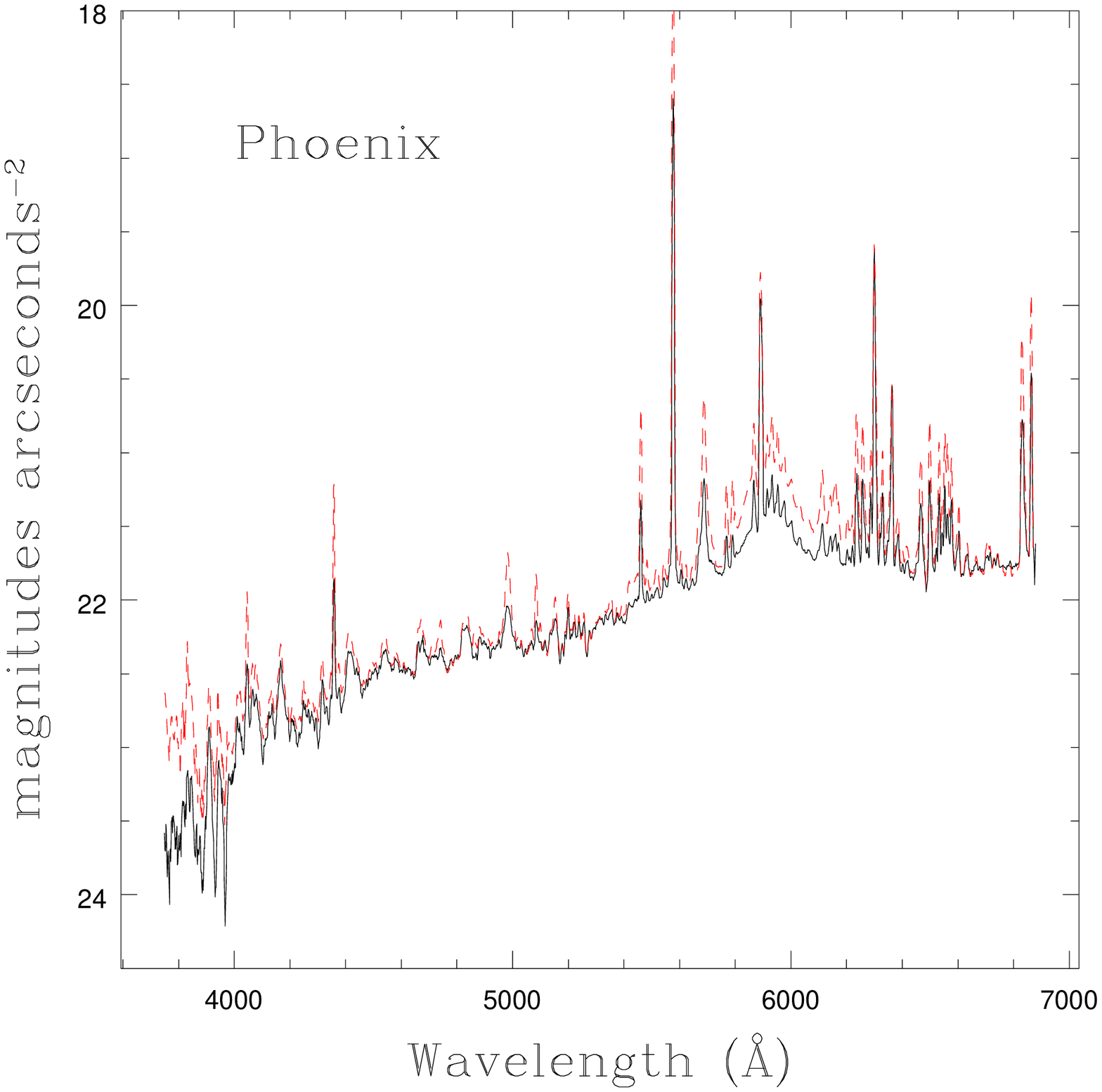}
\plotone{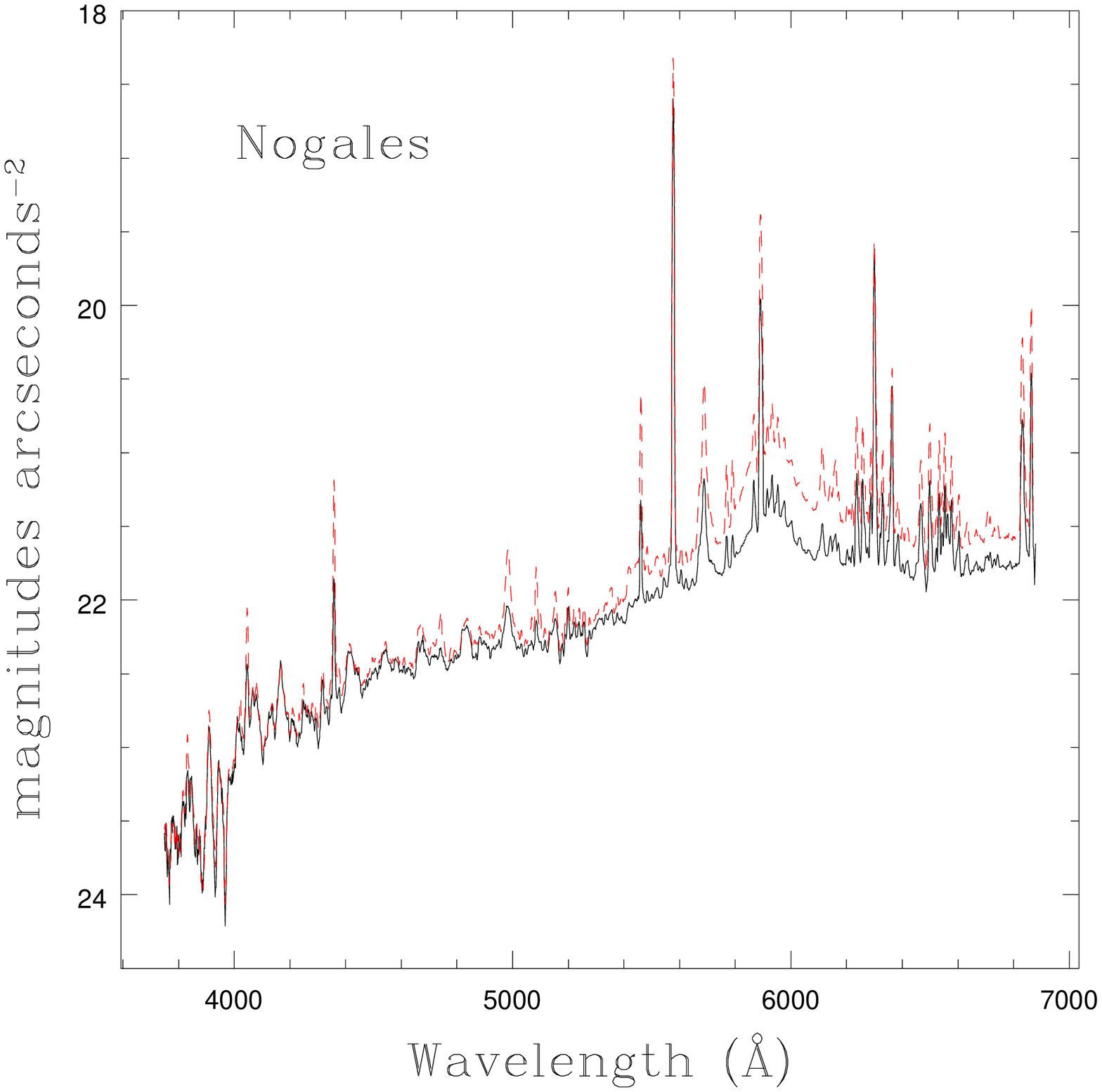}
\plotone{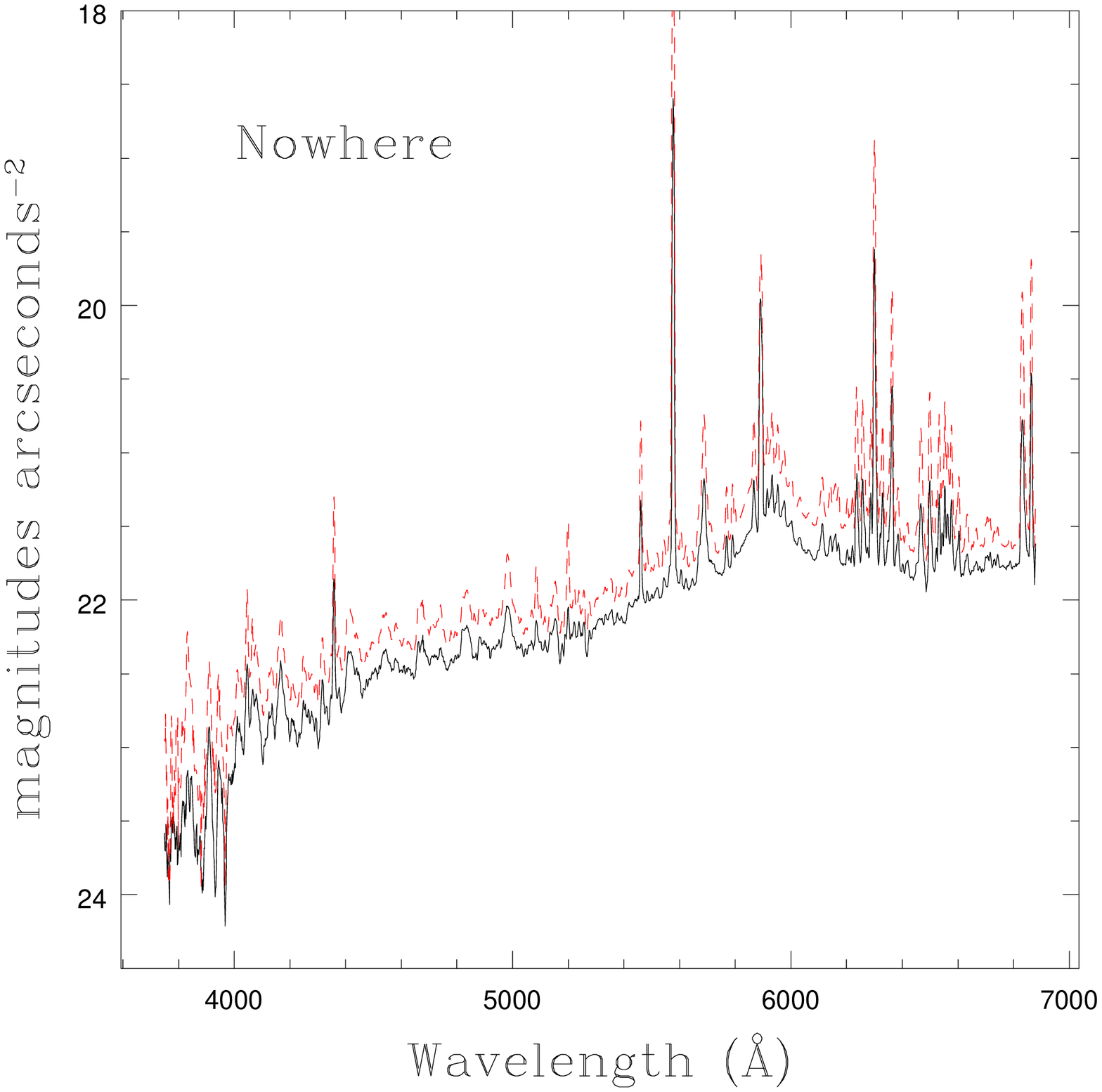}
\caption{\label{fig:ZenComp} Zenith vs.\ Tucson, Phoenix, Nogales and Nowhere for 2009/10. For all four plots, the bottom black line represents the Zenith while the top red dashed line represents:  \emph{(top left)} Tucson, \emph{(top right)} Phoenix, \emph{(bottom left)} Nogales, and \emph{(bottom right)} Nowhere.}
\end{figure}

Comparisons between the average spectra at Zenith and the average spectra at Tucson, Phoenix, Nogales and Nowhere for the 2009/10 data (as shown in Figure~\ref{fig:ZenComp}) reveal just how much the population of surrounding cities affects the sky glow. The most prominent difference is observed when looking at Tucson, as expected, but interestingly there is relatively no difference between the spectra of Nogales and Nowhere. Additionally, according to Table~\ref{tab:CompMags}, the sky towards Phoenix is actually \emph{darker} than the sky towards both Nowhere and Nogales. This suggests that most of the light pollution is being caused by Tucson, even when looking in a completely different direction.

\begin{figure}
\epsscale{1.0}
\plotone{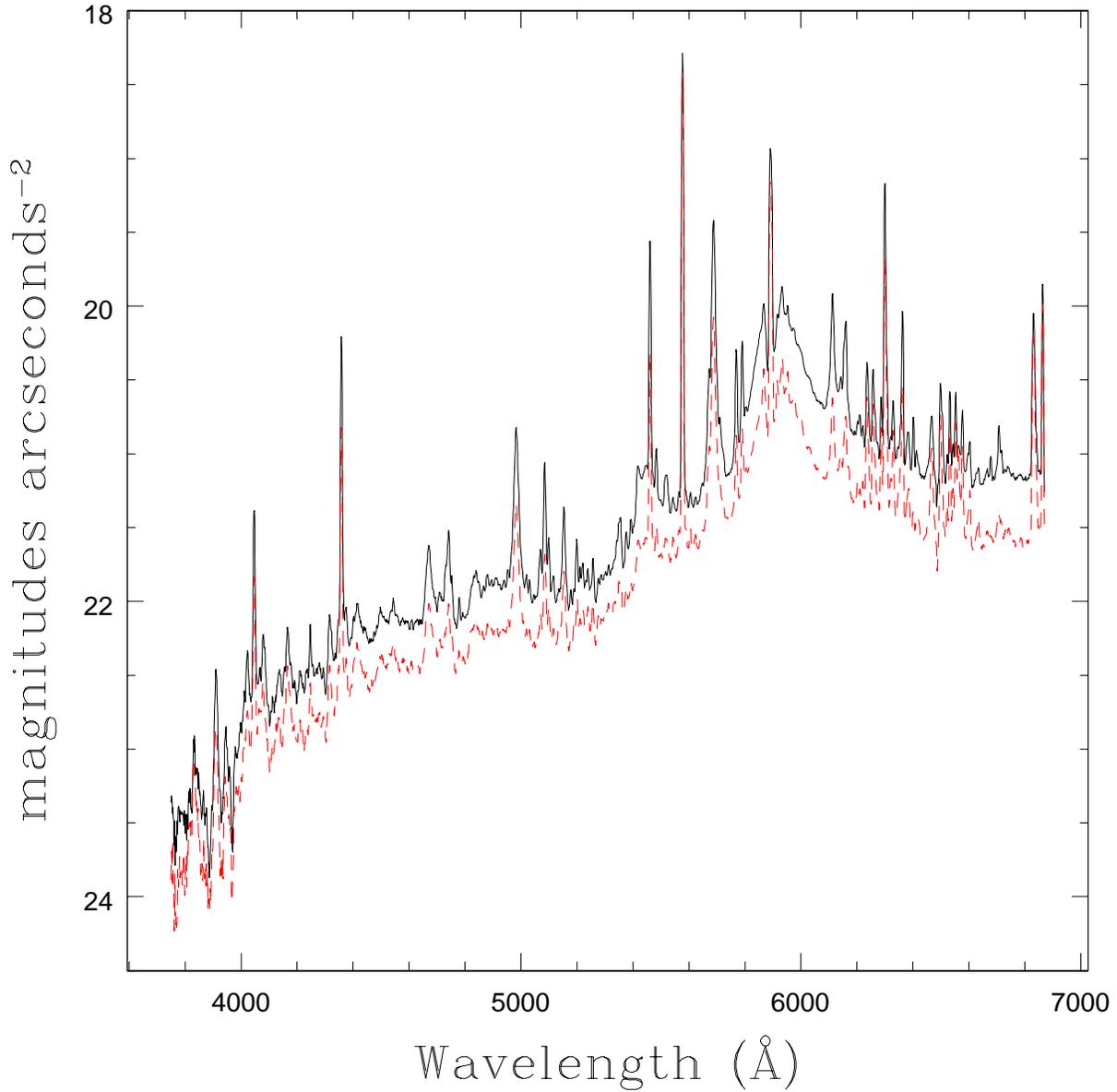}
\caption{\label{fig:TucTime} Changes as the night progressed. Looking toward Tucson in 2010 February, there was a significant change in the sky brightness between the black solid line taken at UT 4:22 ($\sim$3 hours after sunset) and the red dotted line taken at UT 10:20 ($\sim$9 hours after sunset) as homes and businesses turned off their lights. A similar magnitude change throughout the night was found in 2009 October.}
\end{figure}

We also noticed a significant decrease in the sky brightness towards Tucson as the night progressed. Figure~\ref{fig:TucTime} shows the sky spectrum in 2010 February at both three hours after sunset and nine hours after sunset. A comparison of the sky spectrum taken in 2009 October at two hours and eight hours after sunset shows a similar result. Since all of the observations were taken when the sun was greater than 18$^{\circ}$ below the horizon, we believe this change is due to households and businesses switching off their lights rather than the progression of twilight.

We did attempt to quantify the changes seen in the integrated flux of just the region around the Na I D line, which is dominated by high- and low-pressure sodium lamps. The flux in this region is about 1 to 3$\times 10^{-15}$ ergs cm$^{-2}$ s$^{-1}$ \AA$^{-1}$ for the Tucson and Phoenix directions, and smaller for the other directions. The current data show no more flux than the 1988 or 1999 data, and, if anything, is somewhat less. However, the variations within a single night of  observations towards Tucson are more significant than the changes over the past two decades. 

\section{Discussion}
\label{D}
Overall, the sky brightness at Kitt Peak has remained remarkably constant over the past 20 years. While the magnitudes of many artificial sources have increased, this should have little effect on astronomical spectroscopy unless the observer is particularly interested in the Na I D lines.

\begin{figure}
\epsscale{1.0}
\plotone{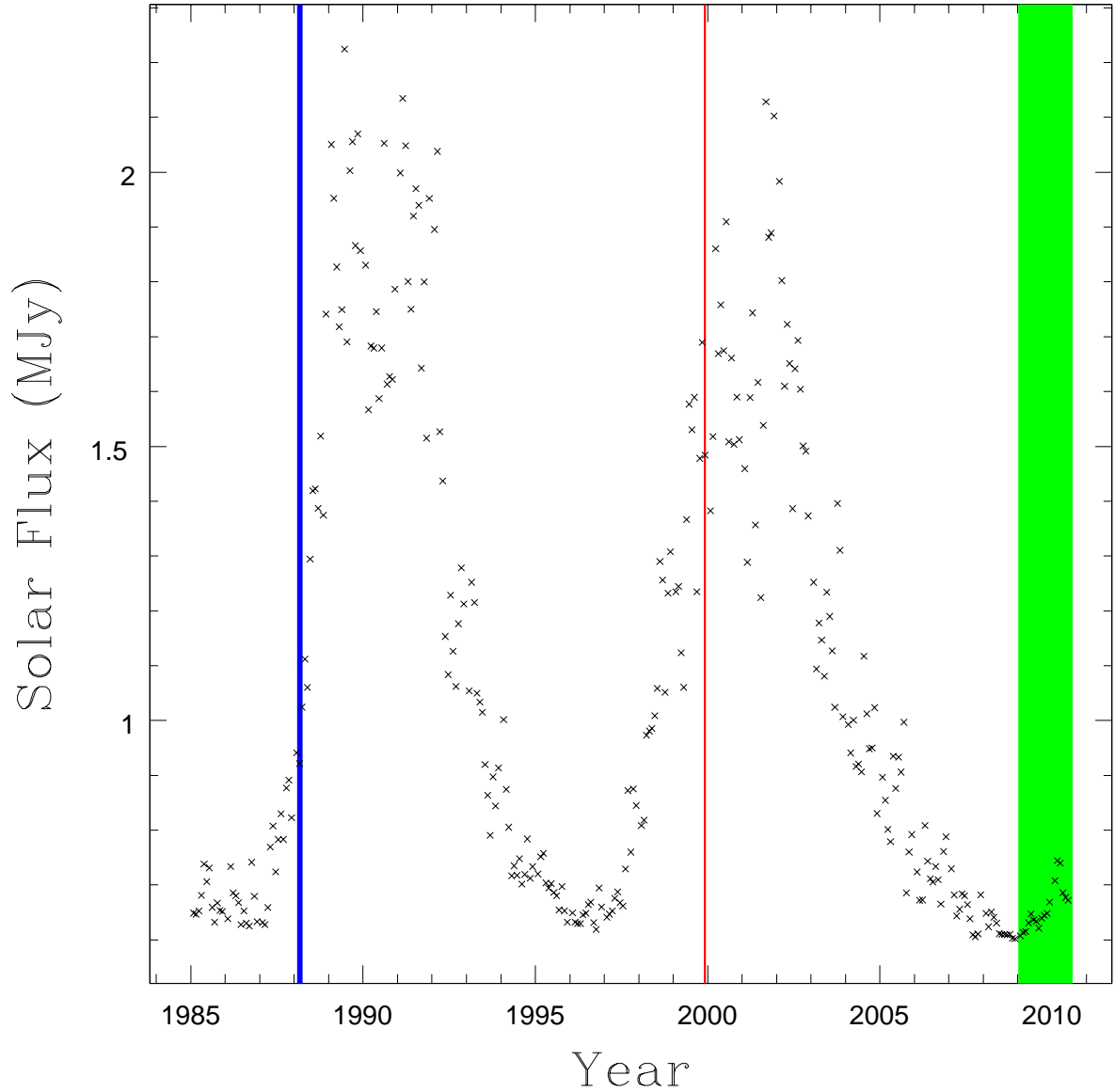}
\caption{\label{fig:sunPlot} Changes in the solar irradiance over the past 30 years. The blue line represents the period of time over which the 1988 data was taken, the red represents the 1999 data and the green represents the 2009/10 data. While each dataset was taken almost a full solar cycle (11 years) apart, the overall solar brightness didn't remain constant. The current data set (2009/10) was taken with the lowest solar activity while the 1999 data was taken with the highest. This data is from the Pentiction-Ottawa solar density flux survey at 2.8 GHz (10.7 cm) available at http://www.ngdc.noaa.gov/stp/solar/flux.html.}
\end{figure}

Other aspects besides artificial sources, such as the solar cycle, also contribute to the sky brightness (Pilachowski et al.\ 1989). As discussed in both Massey et al.\ (1990) and Massey \& Foltz (2000), previous solar maximums occurred in July 1989 and March 2000, corresponding almost perfectly with the first two studies. Even though these investigations have been completed almost one solar cycle (11 years) later, Figure~\ref{fig:sunPlot} shows that the solar flux changed quite drastically between our observations. While the 1999 data was taken at a period close to the solar maximum, the 2009/10 data was taken at a clear solar minimum while the 1988 data was taken at an intermediate stage. This is a likely explanation for why the 1999 data is significantly brighter than the 2009/10 data. The solar cycle should affect the atmospheric lines (such as NI $\lambda$5199, OI $\lambda$5577 and OI $\lambda$6300-64) as well as portions of the blue and red ``pseudo-continuum." However, the NaI or HgI light pollution lines should not be affected.

In an attempt to correct for the effects of the solar cycle on our data, we scaled our $V$ and $B$ magnitudes to the solar minimum using the knowledge that the sky gets brighter by $\sim$0.4 magnitudes between the solar minimum and the solar maximum (Benn \& Ellison 1998, Pedani 2009), and the ``solar phases" (based upon the solar flux) when our observations were taken. These results are shown in Table~\ref{tab:sFlux} and suggest that while the observed sky brightness is comparable to what it was in 1999 (especially considering our 0.04 magnitude error), the sky has brightened at Zenith by approximately 0.1 magnitude since 1988. Similarly, these corrected data suggest that while the sky towards both Tucson and ``nowhere" has increased in brightness by approximately 0.3 magnitudes in $V$ and $B$ since 1988, they are actually $\sim$0.1 magnitude darker than they were in 1999. Once needs to recall that these corrections are approximate, but they help place the measurements on a uniform (if hypothetical) basis. Still, the data is certainly very supportive of the view that lighting ordinances have been an effective tool despite the population growth. 

A comparison of our three datasets (1988, 1999 and 2009/10) highlights the robustness of our current study. In 1988, seven observations were made on a single night. In 1999, ten observations were made on three nearly consecutive nights. In contrast, this study encompasses 30 observations on 6 nights and over a 15 month timespan. Besides the increased baseline, our exposure times were longer by a factor of $\sim$3. Thus, after another decade passes, we hope to carry out a test as rigorous as the one just completed. 

While the arid conditions of southern Arizona failed to stem the region's growth, our results show that the sky brightness hasn't increased proportionally with the population. Over the last two decades, both Pima and Maricopa Counties (which contain Tucson and Phoenix, respectively) have nearly doubled in size and Nogales and Santa Cruz County have experienced similar trends (U.S. Census Bureau: 1990, 2010). However, Pima County's 1974 Lighting ordinance, and its successors in 1987, 2000 and 2005 along with similar regulations imposed by the City of Tucson, have effectively stopped the impact of household, commercial and outdoor lights on the night sky. According to the Pima County Report (Davis et al.\ 2006), light pollution at Kitt Peak has stabilized at 2005 levels and is expected to stay constant until at least 2030. Additionally, in 1974, Turnrose published absolute spectrometry of the Palomar night sky, at a time when Palomar was considered one of the premiere dark observing sites. His $\lambda$4540 flux is equivalent to 22.30 magnitudes arcseconds$^{-2}$ which is comparable to our $\lambda$4550 magnitudes presented in Table~\ref{tab:CompMags}. Thus, Kitt Peak is currently just as dark as Palomar was in the mid-70s.

Astronomy has a major impact on Arizona's state economy -- \$250 million in direct impact every year as well as \$1 billion in infrastructure (Eller College of Management, 2007). Thus, the importance of keeping the skies dark extends far beyond the scope of Astronomy. Lighting ordinances have saved the Kitt Peak night sky thus far, but as the populations of neighboring towns increase and grow closer to the mountain and as towns spring up around I-10, it will become more and more difficult to maintain the current light pollution levels. Still, our results show that Kitt Peak is essentially as dark now as it was in the late 1980s and if the lighting laws toughen as growth increases, there is no reason Kitt Peak can't retain the pristine dark skies that Aden Meinel found so promising. 

\acknowledgements
The authors wish to thank the former Kitt Peak Director, Buell Jannuzi, for encouraging them to obtain a new data set and for being generous with the allocation of observing time to make this study possible. Dr. Jannuzi has been a tireless proponent of keeping the skies over Kitt Peak dark and working within the community to reduce light pollution through education and advocacy. We also thank the Tohono O'odham National for being helpful in their choices for outdoor lighting, the International Dark Sky Association for spreading the word and surrounding governments for their continued willingness to consider the impact of their choices on the conduct of astronomy in southern Arizona. Our study shows that these and other efforts have paid off. But, constant vigilance is essential. We would also like to thank Di Harmer for her strong support and advice, Bill Binkert for helping set up the spectrograph on occasion, Jeff Hall for helping us analyze the effects of the solar cycle on the night sky brightness, the Friends of Lowell Observatory for publication costs, and the anonymous referee for their suggestions that helped improve the paper.

\begin{deluxetable}{l l c c c c c c}
\tablecaption{\label{tab:AllMags} 2009/10 Sky Brightness Measurements (mag arcsec$^{-2}$)}
\tabletypesize{\tiny}
\tablewidth{0pt}
\tablehead{
& &
\multicolumn{2}{c}{Broadband} & &
\multicolumn{3}{c}{Narrowband} \\ \cline{3-4} \cline{6-8}
\colhead{UT} &
\colhead{Azimuth} &
\colhead{$V$} &
\colhead{$B$} &
\colhead{} &
\colhead{$\lambda$4250} &
\colhead{$\lambda$4550} &
\colhead{$\lambda$5150}
}
\startdata
2009 Mar 24.4 & Zenith & 21.96 & 22.85 & & 23.00 & 22.78 & 22.44 \\
2009 Oct 17.2 & & 21.57 & 22.54 & & 22.69 & 22.58 & 22.22 \\
2009 Oct 17.3 & & 21.51 & 22.52 & & 22.69 & 22.53 & 22.16 \\
2010 Feb 14.3 & & 21.91 & 22.75 & & 22.94 & 22.62 & 22.27 \\
2010 Feb 15.3 & & 21.98 & 22.77 & & 22.95 & 22.66 & 22.31 \\
2010 Jun 13.2 & & 22.32 & 23.00 & & 23.16 & 22.98 & 22.69 \\
2010 Jun 13.3 & & 22.32 & 22.85 & & 22.99 & 22.85 & 22.64 \\
2010 Jun 14.3 & & 22.36 & 22.96 & & 23.11 & 22.95 & 22.73 \\ \\
2009 Mar 24.3 & Tucson & 21.40 & 22.51 & & 22.72 & 22.49 & 22.10 \\
2009 Oct 17.1 & & 20.96 & 21.87 & & 22.00 & 21.95 & 21.63 \\
2009 Oct 17.4 & & 20.98 & 22.15 & & 22.35 & 22.19 & 21.77 \\
2010 Feb 14.2 & & 21.16 & 22.37 & & 22.63 & 22.35 & 21.86 \\
2010 Feb 14.4 & & 21.42 & 22.63 & & 22.88 & 22.58 & 22.14 \\
2010 Feb 15.4 & & 21.56 & 22.71 & & 22.95 & 22.65 & 22.22 \\
2010 Jun 13.3 & & 21.63 & 22.42 & & 22.61 & 22.43 & 22.13 \\
2010 Jun 14.3 & & 21.59 & 22.38 & & 22.55 & 22.38 & 22.11 \\ \\
2009 Oct 17.2 & Phoenix & 21.18 & 22.28 & & 22.45 & 22.41 & 22.00 \\
2010 Feb 14.3 & & 21.67 & 22.72 & & 22.93 & 22.66 & 22.23 \\
2010 Feb 15.4 & & 21.78 & 22.86 & & 23.07 & 22.79 & 22.37 \\
2010 Jun 13.3 & & 21.85 & 22.61 & & 22.78 & 22.61 & 22.26 \\
2010 Jun 13.4 & & 21.94 & 22.61 & & 22.78 & 22.63 & 22.35 \\
2010 Jun 14.3 & & 21.76 & 22.51 & & 22.68 & 22.52 & 22.20 \\ \\
2009 Oct 17.2 & Nogales & 21.05 & 22.29 & & 22.47 & 22.35 & 21.91 \\
2010 Feb 14.2 & & 21.73 & 22.73 & & 22.93 & 22.67 & 22.26 \\
2010 Feb 15.3 & & 21.70 & 22.68 & & 22.87 & 22.61 & 22.21 \\ \\
2009 Mar 24.4 & Nowhere & 21.62 & 22.47 & & 22.63 & 22.42 & 22.08 \\
2009 Oct 17.3 & & 21.18 & 22.40 & & 22.56 & 22.46 & 22.03 \\
2010 Feb 14.1 & & 21.78 & 22.78 & & 22.97 & 22.72 & 22.28 \\
2010 Jun 13.2 & & 21.73 & 22.50 & & 22.67 & 22.46 & 22.10 \\
2010 Jun 14.2 & & 21.65 & 22.32 & & 22.48 & 22.31 & 21.99 
\enddata
\end{deluxetable}

\begin{deluxetable}{c l c c c c c c}
\tablecaption{\label{tab:CompMags}Sky Brightness Measurements Over the Past 20 Years (mag arcsec$^{-2}$)}
\tablewidth{0pt}
\tablehead{
& &
\multicolumn{2}{c}{Broadband} & &
\multicolumn{3}{c}{Narrowband} \\ \cline{3-4} \cline{6-8}
\colhead{Year} &
\colhead{Azimuth} &
\colhead{$V$} &
\colhead{$B$} &
\colhead{} &
\colhead{$\lambda$4250} &
\colhead{$\lambda$4550} &
\colhead{$\lambda$5150}
}
\startdata
1988 & Zenith (avg.\ of 2) & 21.95 & 22.84 & & 23.03 & 22.73 & 22.32 \\
     & Tucson (avg.\ of 2) & 21.63 & 22.71 & & 22.93 & 22.66 & 22.21 \\
     & Phoenix & 21.73 & 22.79 & & 23.02 & 22.75 & 22.29 \\
     & Nowhere (avg.\ of 2) & 21.84 & 22.72 & & 22.91 & 22.66 & 22.27 \\
\\
1999 & Zenith (avg.\ of 4) & 21.72 & 22.67 & & 22.82 & 22.64 & 22.22 \\
     & Tucson (avg.\ of 2) & 21.14 & 22.20 & & 22.38 & 22.20 & 21.76 \\
     & Phoenix (avg.\ of 2) & 21.34 & 22.42 & & 22.56 & 22.43 & 21.96 \\
     & Nogales (avg.\ of 2) & 21.34 & 22.39 & & 22.56 & 22.40 & 21.93 \\
\\
 2009 & Zenith (avg.\ of 3) & 21.95 & 22.79 & & 22.96 & 22.69 & 22.34\\
     & Tucson (avg.\ of 6) & 21.46 & 22.50 & & 22.72 & 22.48 & 22.09\\
     & Phoenix (avg.\ of 5) & 21.80 & 22.66 & & 22.85 & 22.64 & 22.28\\
     & Nogales (avg.\ of 2) & 21.72 & 22.71 & & 22.71 & 22.64 & 22.24\\
     & Nowhere (avg.\ of 4) & 21.70 & 22.52 & & 22.69 & 22.48& 22.11\\
\enddata
\end{deluxetable}

\begin{deluxetable}{c l c c c c c c}
\tablecaption{\label{tab:sFlux} The Impact of Solar Flux on Sky Brightness\tablenotemark{*}}
\tablewidth{0pt}
\tablehead{
\colhead{Year} &
\colhead{Azimuth} &
\colhead{$V$} &
\colhead{$B$} &
\colhead{Solar Flux} &
\colhead{Solar Phase} &
\colhead{Adjusted} &
\colhead{Adjusted} \\
& & & &
\colhead{(MJy)}
& &
\colhead{$V$} &
\colhead{$B$}
}
\startdata
1988 & Zenith & 21.95 & 22.84 & 1.23 & 0.29 & 22.07 & 22.96\\
     & Tucson & 21.63 & 22.71 & 0.94 & & 21.75 & 22.83\\
     & Phoenix & 21.73 & 22.79 & 0.94 & & 21.85 & 22.91\\
     & Nowhere & 21.84 & 22.72 & 0.94 & & 21.96 & 22.84\\
\\
1999 & Zenith & 21.72 & 22.67 & 1.73 & 0.57 & 21.95 & 22.90\\
     & Tucson & 21.14 & 22.20 & 1.73 & & 21.37 & 22.43 \\
     & Phoenix & 21.34 & 22.42 & 1.73 & & 21.57 & 22.65\\
     & Nogales & 21.34 & 22.39 & 1.73 & & 21.57 & 22.62\\
\\
 2009 & Zenith & 21.95 & 22.79 & 0.72 & 0.00 & 21.95 & 22.79\\
     & Tucson & 21.46 & 22.50 & 0.70 & & 21.46 & 22.50\\
     & Phoenix & 21.80 & 22.66 & 0.69 & & 21.80 & 22.66\\
     & Nogales & 21.72 & 22.71 & 0.76 & & 21.72 & 22.71\\
     & Nowhere & 21.70 & 22.52 & 0.68 & & 21.70 & 22.52\\
\enddata
\tablenotetext{*}{The solar fluxes and phases were averaged over the months we observed. Monthly apparent solar flux observation data can be found at: ftp://ftp.geolab.nrcan.gc.ca/data/solar\_flux/month\_averages/maver.txt. These monthly averages were then multiplied by a factor of 0.9 to compensate for uncertainties in antenna gain and in waves reflected from the ground.}

\end{deluxetable}

\end{document}